\documentclass[prl,twocolumn,showkeys]{revtex4}
\usepackage{amsmath}
\usepackage{amssymb}
\usepackage{graphicx}
\usepackage{bm}

\pdfoutput=1
\input{tcilatex}

\begin{document}

\title{Observation of asymmetric spectrum broadening induced by silver
nanoparticles in a heavy-metal oxide glass}
\author{N. Zhavoronkov$^{1}$, R. Driben$^{2,3^{\ast }}$, B. A. Bregadiolli$%
^{4}$, M. Nalin$^{4}$ and B. A. Malomed$^{5}$}
\affiliation{$^1$Max-Born-Institut for Nonlinear Optics and Short Pulse Spectroscopy,
Max-Born-Str. 2a, D-12489 Berlin, Germany}
\affiliation{$^2$Institut fur Physik, Universitat Rostock, Universitatsplatz 3, 18055
Rostock, Germany}
\affiliation{$^3$Jerusalem College of Engineering, Ramat Beit HaKerem, POB 3566,
Jerusalem, 91035, Israel}
\affiliation{$^4$LAVIE - Departament of Chemistry - UFSCar, Rodovia Washington Luiz Km
235, 13565-905, S\~{a}o Carlos, SP, Brazil}
\affiliation{$^5$Department of Physical Electronics, School of Electrical Engineering,
Faculty of Engineering, Tel Aviv University, Tel Aviv 69978, Israel}
\date{\today}

\begin{abstract}
We demonstrate experimentally and support by a theoretical analysis an
effect of asymmetric spectrum broadening, which results from doping of
silver nanoparticles into a heavy-glass matrix, 90(0.5WO$_{3}$-0.3SbPO$_{4}$%
-0.2PbO)-10AgCl. The strong dispersion of the effective nonlinear
coefficient of the composite significantly influences the spectral
broadening via the self-phase modulation, and leads to a blue upshift of the
spectrum. Further extension of the spectrum towards shorter wavelengths is
suppressed by a growing loss caused by the plasmon resonance in the silver
particles. The red-edge spectral broadening is dominated by the stimulated
Raman Scattering.
\end{abstract}

\maketitle

%\twocolumn[

%\address{$^*$Corresponding author: radik@eng.tau.ac.il}

%\pacs{42.65.Re, 33.80.-b}

%\ocis{320.6629, 160.4236}

%] %% activate for two-column option

%\section{Introduction}

Dielectric composites containing metal nanoparticles (MNPs) have been
intensively studied for the last two decades (see, e.g., Refs. \cite%
{Panorama, Schmid, Denisyuk}), as they play a central role in the rapidly
growing fields of nano-optics and plasmonics. The interaction between light
and MNPs is dominated by charge-density oscillations on the closed surface
of the particles, alias localized plasmon resonances, and leads to a strong
field enhancement in the MNP's near field. To mention just a few
applications of such systems, localized plasmons allow greatly increased
signal strengths in the Raman spectroscopy and surface spectroscopy,
enabling the detection of single molecules \cite{Nie}, significant
enhancement of the emission rate of fluorescent molecules and quantum dots 
\cite{K?hn}, and the direct generation of high harmonics by nJ pulses
produced by a laser oscillator, without the use of amplifiers \cite{Kim}.
The property of MNPs which is most important for applications to nonlinear
optics is their large intrinsic nonlinearity coefficient. Far from the
plasmon resonance, the third-order susceptibility of silver MNPs in the
nanoparticle composite is about seven orders of magnitude higher than in the
host material \cite{Edilson, Ganeev, Rativa, Kelly, Kohlgraf-Owens}. Since
the local electric field in MNPs is enhanced at the plasmonic resonance, the
nonlinear response of the metal can be additionally significantly amplified
as the wavelength shifts towards the resonance.

The generation of coherent white light (or supercontinuum generation) in
microstructured fibers by fs and ps pulses has attracted a great deal of
attention and found many applications (see review \cite{Dudley}). Recently,
much effort has been devoted to decrease the power or energy threshold
necessary to generate octave-spanning spectra by reducing the relevant
diameters to the submicron scale \cite{Foster} or using nanowires made of
materials with high nonlinearities \cite{Yeom}. The reduction of the power
threshold is particularly important for the realization of cost-effective
and compact SC sources. An alternative approach to lowering the threshold
condition for nonlinear processes is the utilization of composites
containing MNPs. Previously, white-light generation in nanometer-scale
antennas related to photoluminescence \cite{Muhlschlegel}, and in colloidal
MNP\ systems synthesized by means of ablation \cite{Besner} were reported.
Additionally, a relatively small spectral broadening in aqueous colloids
with very low filling factor was observed \cite{Wang}. In recent works, a
mechanism of the wavelength-dependent nonlinearity in aqueous colloids \cite%
{Wang, Driben3}, and in silica glasses volume-doped by silver MNPs, was
studied, and it was predicted that it can lead to the generation of
supercontinuum \cite{Wang, Driben} and solitons \cite{Driben2}. In the
present Letter we produce an experimental evidence of \emph{asymmetric}
spectral broadening due to the presence of nanoparticles in composites based
on WO$_{3}$ host glass, and support these observations by extended numerical
calculations.

The synthesis of the glasses was carried out by melting 99\%-grade-purity
materials WO$_{3}$, PbO, AgCl from Aldrich, and SbPO$_{4}$ prepared as in
Ref. \cite{nal} in a platinum crucible kept in an electrical furnace during
30 minutes at 1100$^{\text{o}}$C. After homogenization, the melt was cooled
to 950$^{\text{o}}$C, left at that temperature for 5 minutes, and then cast
into a brass mold, pre-heated to 400$^{\text{o}}$C. Glass pieces were left
for annealing at that temperature during four hours to reduce thermal
stresses, and then the furnace was turned off and cooled to the room
temperature under its natural cooling rate. Silver MNPs were implanted by
heat treatment of the glass sample under normal pressure for two hours at 450%
$^{\text{o}}$C.

\begin{figure}[tbph]
\centerline{\includegraphics[width=7.9cm]{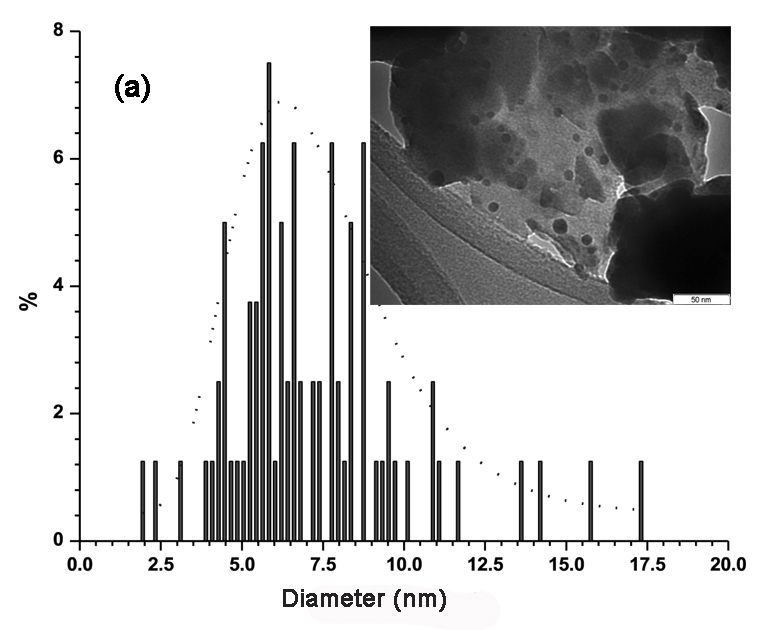}}%%
\centerline{\includegraphics[width=7.9cm]{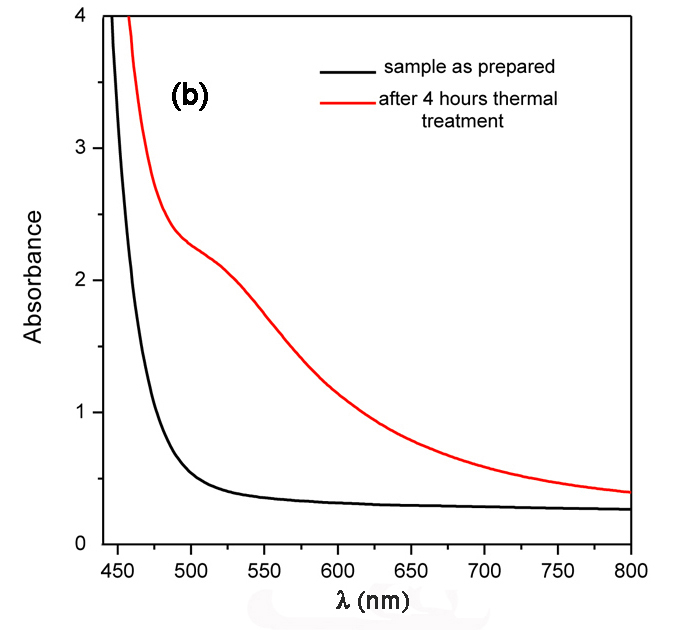}}%%
\caption{(Color online) (a) Size distribution silver nanoparticles
determined by transmission electron microscopy (TEM). A quasi-spherical
geometry as demonstrated in TEM image (inset). (b) The absorption spectrum
of the annealed sample.}
\label{Fig.1}
\end{figure}

The size distribution and shape of silver MNPs have been determined by means
of the transmission electron microscopy (TEM). The particles demonstrate a
quasi-spherical shape, as it is can be observed in the TEM image in the
inset to Fig. \ref{Fig.1}(a). The result of the MNP analysis for the size
distribution is presented in Fig. \ref{Fig.1}(a), giving the mean diameter
around 6 nm. The absorption spectrum of the annealed sample in Fig.1 (b)
demonstrates a new absorption shoulder at $\symbol{126}515$ nm, in
comparison to the ``as prepared" glass composition, which is attributed to
the surface plasmon resonance of silver. The strong red-shift of the
resonance in this case is accounted for by the high linear refractive index
of glass matrix, $n\approx 2.11$.

A Ti:sapphire-based laser system, generating 30-fs pulses centered at 800 nm
with up to 3 mJ energy, was used in the experiment. The spatial profile of
the output beam was measured to be Gaussian with quality parameter $M^{2}$
close to $1$. The diameter of the laser beam was reduced with a use of a 1:3
telescope to 5 mm at the $1/e$ level. The beam's power was attenuated by a
factor of $10$ with the help of a rotating optical chopper, to avoid thermal
effects in the samples. The 2 mm thick WO$_{3}$ glass samples, both doped
with the silver MNPs and pure ones, were placed between a pair of mirrors,
introducing dispersion of $-40$ fs$^{2}$ per bounce, to compensate the
material dispersion of the glass matrix and prevent the distortion of the
pulse in the temporal domain. The mirror-pair was also used to arrange
additional passes of the pulses through the samples, by adjusting the
relative position of the mirrors. It was not reasonable to perform more that
four passes through the silver-doped sample because of a relative low
single-pass transmissivity of $70\%$, which is mostly determined by high
losses due to the reflection from uncoated glass surfaces, with the
respective refractive index $n>2$, and by imperfections of the surfaces. The
spectrum of the transmitted beam was measured using an intensity-calibrated
spectrometer (AvaSpec-2048 190-1100 nm), after attenuating the Fresnel
reflection from the fused silica edge.

If both the average distance between the MNPs and their size are much
smaller than the wavelength of light, the effective-medium description,
based on the Maxwell-Garnett model, can be applied to spherical particles
with uncorrelated positions, up to relatively large values of the filling
factor, $f\lesssim 0.2$. In this model the effective dielectric constant of
the composite is given by $\epsilon _{\mathrm{eff}}=\epsilon _{h}(1+2\sigma
f)/(1-\sigma f)$ with $\sigma \equiv (\epsilon _{i}-\epsilon _{h})/(\epsilon
_{i}+2\epsilon _{h})$. Here $\epsilon _{h}$ and $\epsilon _{i}$ are the
dielectric permittivities of the host and of the MNPs, respectively. The
dielectric permittivity of WO$_{3}$ glass is $\epsilon _{h}$ $=4.2$ at the
central operating wavelength of $800$ nm. For the dielectric permittivity of
the silver MNPs we use the known polynomial approximation \cite{Tanabe}: $%
\epsilon _{i}$ $=\epsilon _{R}+i\epsilon _{I}$, with%
\begin{gather}
\epsilon _{R}=-2.037181\times 10^{-17}\lambda ^{6}+1.183540\times
10^{-13}\lambda ^{5}  \notag \\
-2.537882\cdot 10^{-10}\lambda ^{4}+2.430043\cdot 10^{-7}\lambda ^{3}  \notag
\\
-1.420089\cdot 10^{-4}\lambda ^{2}+8.990214\cdot 10^{-4}\lambda +8.526028, 
\notag
\end{gather}%
\begin{gather}
\epsilon _{I}=-2.327098\cdot 10^{-17}\lambda ^{6}+1.471828\cdot
10^{-13}\lambda ^{5}  \notag \\
-3.635520\cdot 10^{-10}\lambda ^{4}+4.530857\cdot 10^{-7}\lambda ^{3}  \notag
\\
-2.946733\cdot 10^{-4}\lambda ^{2}+9.56229\cdot 10^{-2}\lambda -11.49465. 
\notag
\end{gather}

For the filling factor of the sample used in\ the experiment, $f=1.2\times
10^{-4}$, the group-velocity-dispersion coefficient at 800 nm is $\beta
^{\prime \prime }(\omega )\equiv \partial ^{2}k_{\mathrm{eff}}/\partial
\omega ^{2}=0.0063$ fs$^{\text{2}}/\mathrm{\mu }$m, where $k_{\mathrm{eff}%
}=n_{\mathrm{eff}}(\omega )\omega /c$ and $n_{\mathrm{eff}}(\omega )=\sqrt{%
\epsilon _{\mathrm{eff}}}$, and the linear-loss coefficient is $\alpha =%
\mathrm{Im}\{k_{\mathrm{eff}}(\omega )\}=1.76\times 10^{-4}~\mathrm{\mu }$m$%
^{-1}$, $c$ being the light velocity in vacuum.

\begin{figure}[tbph]
\centerline{\includegraphics[width=7.9cm]{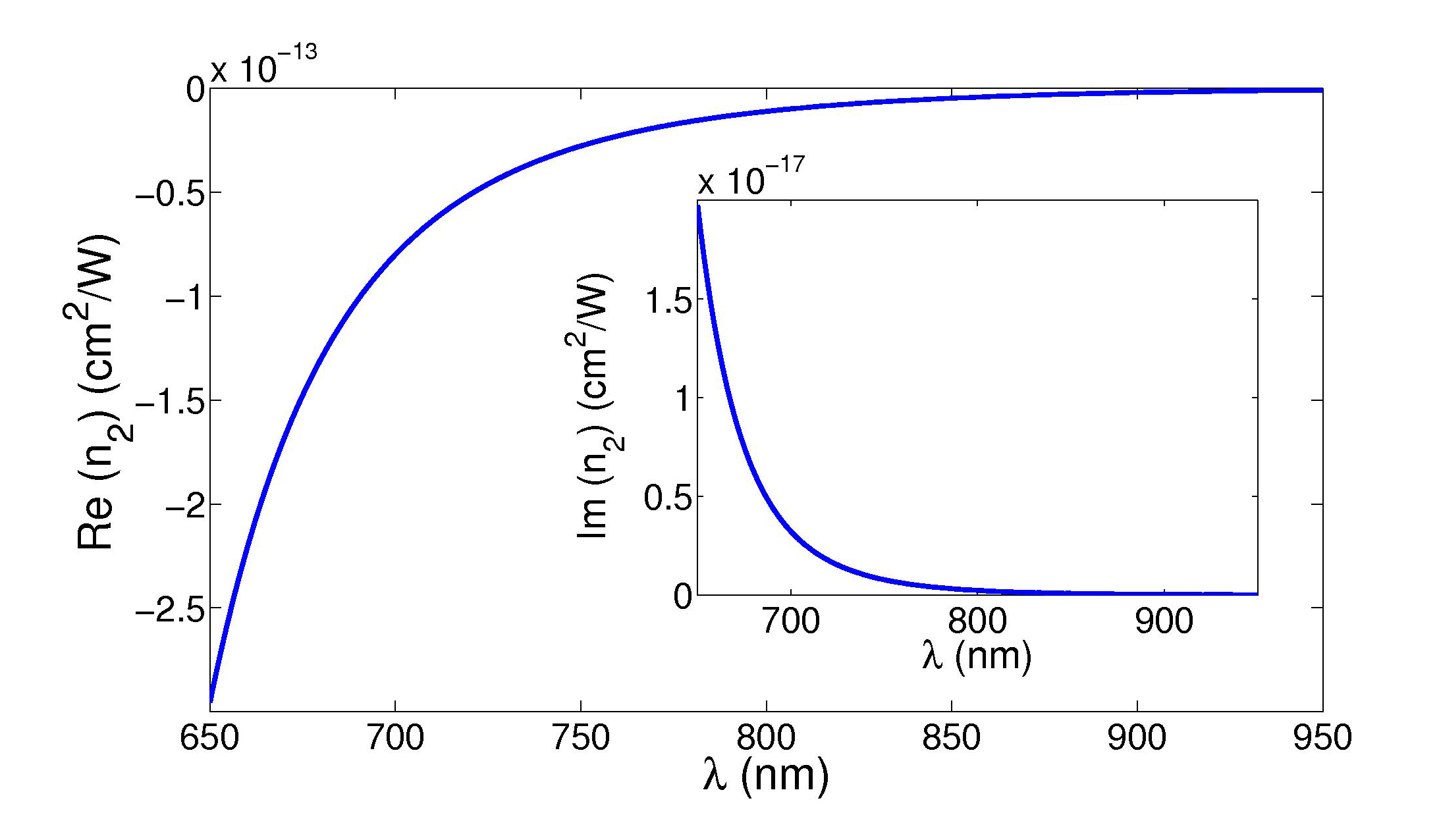}}%%
\caption{(Color online) Nonlinear properties of the heavy glass composite
with silver nanoparticles. Shown are real and imaginary parts of the
nonlinear coefficient, $n_{2}(\protect\omega )=3\protect\chi _{\mathrm{eff}%
}^{(3)}/(4c\protect\epsilon _{\mathrm{eff}}\protect\epsilon _{0}).$}
\label{Fig.2}
\end{figure}

Both the dispersion and losses strongly increase as the plasmon resonance is
approached (see particular curves in Ref. \cite{Driben}). For the filling
factor $f=1.2\times 10^{-4}$ the dispersion length is few cm, while the
effective propagation length and nonlinearity length are few mm, hence the
dispersion effect is negligible. For spherical MNPs, the effective
third-order susceptibility is \cite{Sipe}

\begin{gather}
\chi _{\mathrm{eff}}^{(3)}=f\frac{\chi _{i}}{|P|^{2}P^{2}}+  \notag \\
+\frac{\chi _{h}\left( 1-f+1.6f\sigma ^{2}|\sigma |^{2}+1.2f\sigma |\sigma
|^{2}\right) }{(1-f\sigma )^{2}\left\vert 1-f\sigma \right\vert ^{2}}  \notag
\\
+\frac{\chi _{h}(0.4f\sigma ^{3}+3.6f|\sigma |^{2}+3.6f\sigma ^{2})}{%
(1-f\sigma )^{2}\left\vert 1-f\sigma \right\vert ^{2}},  \label{nonl}
\end{gather}%
with $P~\equiv $ $(1-f\sigma )(\epsilon _{i}+2\epsilon _{h})/3\epsilon _{h}$%
, and $\sigma =(\epsilon _{i}-\epsilon _{h})/(\epsilon _{i}+2\epsilon _{h})$%
, as above. The susceptibility of the host glass is taken to be $\chi
_{h}=1.1\cdot 10^{-21}$m$^{2}/$V$^{2}$, which is close to values reported
for heavy glasses of the same class \cite{Santos}, which are typically five
times higher than the susceptibility of silica glasses. We use the
experimentally measured value $\chi _{\mathrm{Ag}}^{(3)}=(-6.3+1.9i)\cdot
10^{-16}$ m$^{2}/$V$^{2}$ \cite{Edilson} for the susceptibility of silver
nanoparticles. The nonlinear susceptibility of the composite, calculated at
the central operation wavelength of $800$ nm, is $\chi _{\mathrm{eff}%
}^{(3)}=(-1.8+0.36i)\cdot 10$ $^{-20}~$m$^{2}/$V$^{2}$. The strong intrinsic
nonlinearity of the MNPs tends to make the effective nonlinearity of the
composite self-defocusing. Even far from the plasmon resonance, the
magnitude and sign of\textbf{\ }$n_{2}(\omega )$ in the composite strongly
depend on the frequency, as shown in Fig. \ref{Fig.2}. The sign changes to
the self-focusing around 1020 nm, similar to what was reported in Ref. \cite%
{Driben}, which is far away from the spectral range discussed below.

To simulate the propagation of the optical pulse through the composite, we
use the generalized nonlinear Schr\"{o}dinger equation \cite{Agrawal} for
amplitude $A$ of the electric field. This model takes into account the
frequency dependence of the nonlinearity, dispersion of all orders, and the
temporal delay caused by the stimulated Raman Scattering (SRS): 
\begin{gather}
\frac{\partial A}{\partial z}=\sum\limits_{m\geq 2}\frac{i^{m+1}\beta _{m}}{%
m!}\frac{\partial ^{m}A}{\partial \tau ^{m}}-\frac{\alpha }{2}A  \notag \\
+i\gamma (\omega )\left( |A|^{2}-T_{R}\frac{\partial |A|^{2}}{\partial \tau }%
\right) A.  \label{main}
\end{gather}%
Here $z$ is the propagation distance, $\tau $ the time in the reference
frame traveling along with the pump light, $\beta _{m}$ the $m$--th order
dispersion coefficient at the central frequency, and $\alpha $ the linear
loss. The nonlinear coefficient is given by $\gamma (\omega )=n_{2}(\omega
)\omega /(cA_{\mathrm{eff}}),$ where $n_{2}$ is the nonlinear refractive
index of the composite and $A_{\mathrm{eff}}$ the effective mode area. Since
the value of the SRS temporal delay in our glass species was not found in
literature, it was chosen by fitting to empirical data, $T_{R}=0.32$.

\begin{figure}[tbph]
\centerline{\includegraphics[width=7.9cm]{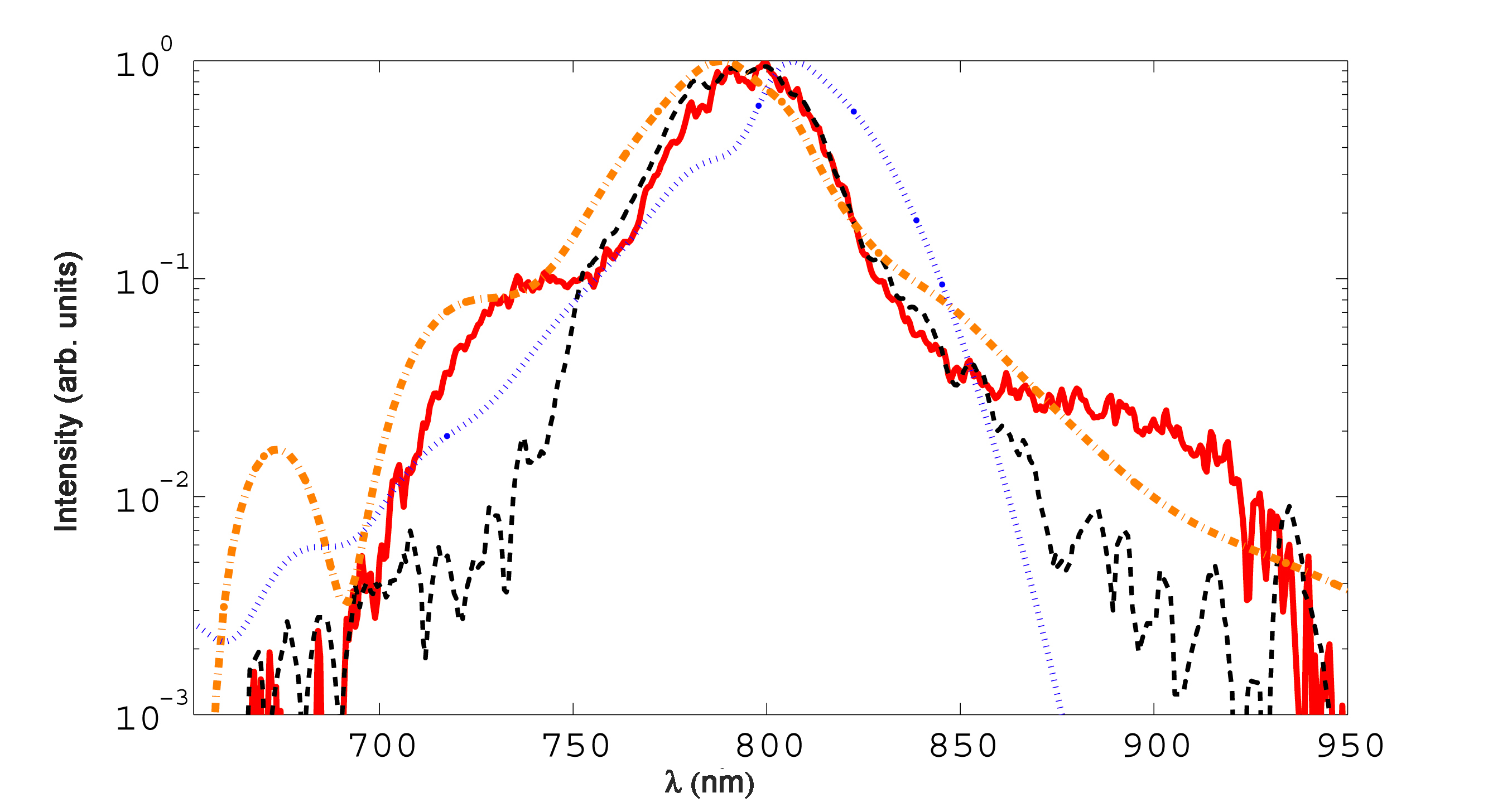}}%%
\caption{(Color online) The experimentally measured spectral broadening
after the $4$ mm-long propagation through the composite sample with and
without the nanoparticles (the solid red and dashed black curves,
respectively). The numerically evaluated spectrum, with the Raman term taken
into regard, is shown by the dot-dashed orange curve, and the one without
the Raman term by the dotted blue curve.}
\label{Fig.3}
\end{figure}

The experimentally measured spectrum in the case of the double propagation ($%
z=4$ mm) of the optical pulse with energy of $1.57$ mJ through the composite
containing silver MNPs is presented in Fig. \ref{Fig.3} by the solid red
curve, together with the spectrum of the same pulse propagated through the
host glass, but without nanoparticles (dashed blue curve). Additional curves
demonstrate the numerically calculated spectra, including (dot-dashed orange
curve) and excluding (dotted blue curve) the Raman term in Eq. (\ref{main}). 

In Fig. \ref{Fig.3}, one can clearly observe an asymmetry in the spectral
broadening. The short-wavelength extension of the spectrum originates from
the increase of the magnitude of the nonlinearity as the wavelengths
decrease toward the plasmon resonance, in accordance with the theoretical
prediction \cite{Driben}. The additional long-wave lower shoulder of the
spectrum arises from the SRS shift. Comparing the two numerical curves, one
concludes that the inclusion of the SRS effect, which was not considered in
the previous analysis reported in Ref. \cite{Driben}, improves the
qualitative and quantitative agreement with the experimental results. To
study the dependence of the spectral broadening on the injected energy,
pulses with different values of the peak power were subjected to the double
propagation through the sample. The more powerful pulses were injected, the
more manifestations of the nonlinear effects in the spectral broadening was
observed, as demonstrated in Fig. 4. However, the use of 2mJ pulses leads to
optical damage of the sample. Additional experiments were performed to
establish an optimal propagation length for the pulse broadening. After the
single propagation through the glass composite, some initial broadening was
observed. The spectral broadening reaches its saturation after the double
propagation through the sample ($z=4$ mm), with very little change observed
after additional propagation cycles.

\begin{figure}[tbph]
\centerline{\includegraphics[width=7.9cm]{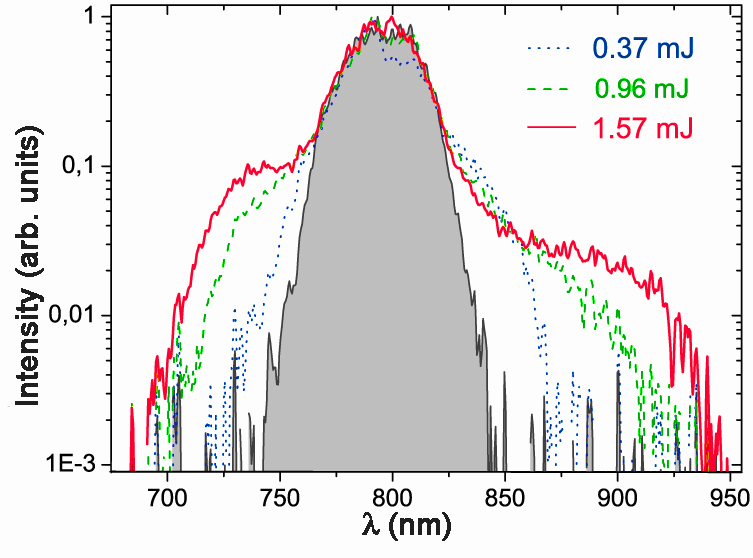}}%%
\caption{(Color online) The spectral broadening after the $4$ mm long
propagation through the composite sample with the nanoparticles, for
different energies of the input pulse. The shaded spectrum represents the
propagation through the host glass without nanoparticles.}
\label{Fig.4}
\end{figure}

In conclusion, we have demonstrated the asymmetric spectral broadening of
the transmitted light pulses due to the presence of silver nanoparticles in
WO$_{3}$-SbPO$_{4}$-PbO host glasses. The composite exhibits the strong
wavelength dependence of the effective nonlinear coefficient, which
dominates the shape of the spectral broadening via the self-phase-modulation
and leads to the blue spectral upshift. The red wing of the spectrum is
induced by the Raman self-frequency shift.

M.S. thank J. M. Caiut for the help with analysis of TEM images.

\newpage

\section*{Figures captions}

Fig.1 (a)Size distribution silver nanoparticles determined by transmission
electron microscopy (TEM). A quasi-spherical geometry as demonstrated in TEM
image (inset). (b)The absorption spectrum of the annealed sample.

Fig.2 Nonlinear properties of the heavy glass composite with silver
nanoparticles. Shown are real and imaginary parts of the nonlinear
coefficient, $n_{2}(\omega )=3\chi _{\mathrm{eff}}^{(3)}/(4c\epsilon _{%
\mathrm{eff}}\epsilon _{0}).$

Fig.3 The experimentally measured spectral broadening after the $4$ mm
propagation through the composite sample with and without the nanoparticles
(the solid red and dashed black curves, respectively). The numerically
evaluated spectrum, with the Raman term taken into regard, is shown by the
dot-dashed orange curve, and the one without the Raman term by the dotted
blue curve.

Fig.4 The spectral broadening after the $4$ mm propagation through the
composite sample with the nanoparticles for different energies of the input
pulse. The shadded spectrum represents the case of propagation through the
host glass without nanoparticles.

\end{document}